\documentclass{tlp}
\usepackage{amssymb}
\usepackage{amsmath}
\usepackage{stmaryrd}
\usepackage{verbatim}
\usepackage{url}
\usepackage{amssymb}
\usepackage{pslatex}

\renewcommand{\emptyset}{\mathord{\varnothing}}

\newcommand{\state}[2]{\langle #1\| #2\rangle}

\newlength{\Blweite}
\def\i4{\settowidth{\Blweite}{1234}\makebox[\Blweite]{}}

{\end{tabbing}%
\end{minipage}%
}

\newsavebox{\savepar}

\newtheorem{definition}{Definition}
\newtheorem{example}{Example}

\newtheorem{theorem}{Theorem}

\title[Recurrence with affine level mappings is P-time decidable for CLP(${\mathbb R}$)]{Recurrence with affine level mappings is P-time decidable for CLP(${\mathbb R}$)\\ {\em Technical note}}
\author[Fred Mesnard \and Alexander Serebrenik]
{FRED MESNARD\\
IREMIA, Universit{\'e} de la R{\'e}union, France\\
\email{Frederic.Mesnard@univ-reunion.fr}
\and 
ALEXANDER SEREBRENIK\\
Laboratory for Quality Software (LaQuSo), T.U. Eindhoven, The Netherlands\\
         \email{A.Serebrenik@tue.nl}}
\submitted{2 May 2006}
\revised{13 October 2006}
\accepted{4 January 2007}

\begin{document}
\newcommand{\eat}[1]{}
\newcommand\othrpearl{}
\maketitle
\label{firstpage}

\begin{abstract} 
In this paper we introduce a class of
constraint logic programs such that their termination
can be proved by using affine level mappings. We show that membership to this 
class is decidable in polynomial time. 
\end{abstract}
\begin{keywords} 
constraint logic  programming -- termination -- decidability
\end{keywords}

\section{Introduction}
\label{Intro}
Termination is well-known to be one of the crucial properties of
software verification. Logic programming, and more generally constraint logic programming
(CLP), with their strong theoretical
basis lend themselves easily to termination analysis as witnessed by
a very intensive research in the area. 

In this paper, which is a
revised version of~\cite{DBLP:conf/lopstr/SerebrenikM04}, we study decidability of 
termination for CLP($\mathbb{C}$) 
programs for a given constraint domain $\mathbb{C}$.
In general, decidability depends on the constraint domain $\mathbb{C}$. On the one hand, Devienne
et al. \shortcite{Devienne:Lebegue:Routier} have established undecidability of termination for 
one-rule binary CLP($\mathbb{H}$) programs, where $\mathbb{H}$ is the domain of
Herbrand terms. On the other hand, Datalog, i.e., logic
programming with no function symbols, provides an example of a constraint
programming language such that termination is decidable. We note that the 
decidability of the related problem of {\em boundedness} for Datalog
queries has been studied, for instance, in~\cite{Afrati:Cosmadakis:Foustoucos,%
Marcinkowski}.
For constraint domains with the undecidable termination property, 
we are interested in subclasses of  
programs such that termination is decidable for these subclasses.
A trivial example is the subclass of non-recursive
programs. 

We organise the paper as follows.
After the preliminary remarks of Section~\ref{section:preliminaries},
in Section~\ref{section:llm} we present our main result.
Section \ref{Related Works}  reviews related  results
before our conclusion.

\section{Preliminaries}
\label{section:preliminaries}

For CLP-related definitions, we follow \cite{JaffarMMS98}.
Extensive introductions to CLP can be found in \cite{Jaffar:Maher,MS98}.
The key notions of CLP are those of an algebra and an associated constraint solver over 
a class of constraints, namely a set of first order formulas 
including the always satisfiable constraint $\mbox{\sl true}$, 
the unsatisfiable constraint $\mbox{\sl false}$, and 
closed under variable renaming, conjunction and existential quantification.
If $c$ is a constraint, we write $\exists c$ for its existential closure.
We consider \emph{ideal} CLP($\mathbb{C}$), i.e., we require the existence of
a constraint solver $\mbox{\sl solv}_{\mathbb{C}}$ mapping 
in finite time each constraint to {\tt true} or {\tt false} such that
if $\mbox{\sl solv}_{\mathbb{C}}(c) =$ {\tt false}
then the constraint $\exists c$ is {\sl false} with respect to  $\mathbb{C}$
 and if $\mbox{\sl solv}_{\mathbb{C}}(c) =$ {\tt true}
then the constraint $\exists c$ is {\sl true} 
with respect to $\mathbb{C}$. 
The associated domain is denoted  $D_{\mathbb{C}}$.
Given a constraint $c$, a {\em solution} of $c$ is a mapping $\theta$ from the set of variables to 
$D_{\mathbb{C}}$ such that $c\theta$ is true with respect to $\mathbb{C}$. 
The set of predicate symbols associated with $\mathbb{C}$ is denoted
$\Pi_{\mathbb{C}}$. 
We are interested in the following domains and languages:
\begin{itemize}
\item $\mathbb{N}$. The predicate symbols are $=$ and $\geq$, the 
function symbols are $0$, $1$, and $+$. 
\item $\mathbb{Q}$ and $\mathbb{R}$. The predicate and function 
symbols are as above.
$\mathbb{Q}^+$ and $\mathbb{R}^+$ restrict $\mathbb{Q}$ and $\mathbb{R}$
to non-negative numbers.
\end{itemize}

Given a CLP($\mathbb{C}$)-program $P$, 
we define $\Pi_P$ as the set of user-defined  predicate symbols
appearing in $P$. We restrict our attention to {\em flat} 
programs, i.e., finite sets of  
rules in a {\em flat} form. So each rule is of form:
either $q_0(\tilde{y_0}) \leftarrow c$ or
$q_0(\tilde{y_0}) \leftarrow c, q_1(\tilde{y_1}), \ldots,q_n(\tilde{y_n})$
where $c$ is a constraint, $q_0, \ldots, q_n \in \Pi_P$, $\tilde{y_0}, \ldots, \tilde{y_n}$ denote
tuples of \emph{distinct} variables,
$\bigcap_{i=0}^n \tilde{y_i} = \emptyset$,
and the set of free 
variables of the constraint $c$ is included in
$\bigcup_{i=0}^n \tilde{y_i}$. 
Flat queries are defined accordingly. 
A {\em binary} program is 
a flat program such that all rules have no more than one user-defined
body subgoal.
The  $\mathbb{C}$-base $B_P^{\mathbb{C}}$ 
is defined as $\{p(d_1,\ldots,d_n)\mid p\in \Pi_P, (d_1,\ldots,d_n)\in 
(D_{\mathbb C})^n\}$.
For a flat query $Q$ of the form $c, A_1,\ldots,A_n$, the set of ground instances of $Q$, 
denoted $\mbox{\sl ground}_{\mathbb C}(Q)$, is the set of conjunctions of the form
$A_1\theta,\ldots,A_n\theta$ where $\theta$ is a solution of $c$.
The notion of groundedness is extended to flat rules and programs.

\begin{example}
\label{example:72}
Consider the following CLP($\mathbb{Q}$) program $P$:
\[ \begin{array}{llll}
r_1 & p(x) & \leftarrow & x=2.\\
r_2 & p(x) & \leftarrow & 0=1.\\
r_3 & p(x) & \leftarrow & 72 \geq x, y = x+1, p(y).
\end{array} \]
This program is a binary program,
$\mbox{\sl ground}_{\mathbb Q}(r_1)$ is $\{p(2)\}$, 
$\mbox{\sl ground}_{\mathbb Q}(r_2)$ is $\emptyset$, 
$\mbox{\sl ground}_{\mathbb Q}(r_3)$ 
is an infinite set that contains, among others, $p(72)\leftarrow
p(73)$ and $p(1/2)\leftarrow p(3/2)$, and 
$\mbox{\sl ground}_{\mathbb Q}(P)=\mbox{\sl ground}_{\mathbb Q}(r_1) \cup 
\mbox{\sl ground}_{\mathbb Q}(r_2) \cup \mbox{\sl ground}_{\mathbb Q}(r_3)$.
Note that ground instances do not contain
any constraint.
\end{example}

We now discuss the operational semantics of CLP-programs
we consider in this paper. A {\em state} of computation is 
a pair $\state{A_1,\ldots,A_n}{c}$.
We further assume that one of
the atoms in $A_1,\ldots,A_n$, say $A_i$, is selected for resolution
by a {\em selection rule}.
The operational semantics can be expressed 
by means of the following rewriting rules:
\begin{itemize}


	\item $\state{A_1,\ldots,A_n}{c}$ rewrites to 
	$\state{\Box}{\mbox{\sl false}}$ if there exists a fresh rule $A'_i
	\leftarrow c', B_1, \ldots, B_m$ in $P$ such that 
	$c\wedge (A_i = A'_i) \wedge c'$ is unsatisfiable;

	\item $\state{A_1,\ldots,A_n}{c}$ rewrites to
$\langle A_1,\ldots,A_{i-1},B_1,\ldots,B_m,A_{i+1},\ldots,A_n || c\wedge A_i = A'_i \wedge c'\rangle$
	if there exists a fresh rule $A'_i
	\leftarrow c', B_1, \ldots, B_m$ in $P$ such that 
	$c\wedge (A_i = A'_i) \wedge c'$ is satisfiable.

\end{itemize}
A {\em derivation} from a state $S_0$ is a finite or infinite sequence of states
$S_0, S_1, \ldots,S_n,Ê\ldots$ such that each $S_i$ can be rewritten
as $S_{i+1}$.
A {\em ground}Ê state is a state $\state{A_1,\ldots,A_n}{\mbox{\sl true}}$
where each $A_i$ belongs to $B_P^{\mathbb{C}}$.
We say that a CLP($\mathbb{C}$) 
program $P$ is {\em terminating} if {\em every} derivation 
starting from {\em any} ground state  
via {\em any}  selection rule 
is finite, under the operational semantics defined above.

To characterize this notion of termination, we use the notion of {\em level mapping}.
A {\em level mapping} for a constraint domain $\mathbb{C}$
is a function $|\cdot |: B_P^\mathbb{C}\rightarrow \mathbb{R}$.
We adapt the idea of recurrence, originally introduced in~\cite{Bezem93}, 
to  CLP:
\begin{definition}
\label{recurrence}
Let $P$ be a flat CLP($\mathbb{C}$) program, 
and $|\cdot |: \mbox{$\mathbb{C}$-{\sl base}}\rightarrow \mathbb{R}$ 
be a level mapping. 
$P$ is called {\em recurrent} with respect to $|\cdot |$ 
if there exists a real number
$\epsilon > 0$ such that,  for every $A\leftarrow B_1, \ldots, B_n\in \mbox{\sl ground}_{\mathbb{C}}(P)$, 
$|A| \in \mathbb{R}^+$, and $|B_i| \in \mathbb{R}^+$, 
$|A| \geq |B_i| + \epsilon$ for all $i$, $1\leq i\leq n$. 
We say that $P$ is {\em recurrent} if there exists a level-mapping 
such that $P$ is recurrent with respect to it.
\end{definition}

Observe that rules of the form $p(\tilde{x})\leftarrow c$ are not 
taken into account by the definition above. 
Moreover, without loss of generality, we may
fix $\epsilon$ to $1$: if $P$ is recurrent in this narrow sense,  $P$
is trivially recurrent with respect to Definition \ref{recurrence}. 
Conversely, since $\epsilon > 0$, we can safely multiply the values of the level mapping by $1/\epsilon$.

\begin{theorem} \cite{Bezem93}
\label{thm:bezem}
$P$ is recurrent if and only if $P$ is terminating.
\end{theorem}

\section{Alm-recurrent programs}
\label{section:llm}
Let us consider 
programs that can be analyzed by means of affine level
mappings. 

\begin{definition}
\label{definition:llm}
A level mapping $\mid\cdot\mid$ is called {\em affine} if
for any $n$-ary predicate symbol $p \in \Pi_P$,
there exist real numbers $\mu_{p,i}$, $0\leq i\leq n$, such that
for any atom $p(e_1,\ldots,e_n)\in B_P^\mathbb{C}$:
$$|p(e_1,\ldots,e_n)| = \mu_{p,0} + \sum_{i=1}^{n} \mu_{p,i} e_i$$
\end{definition}

So for a given atom $p(\tilde{e})$, its affine level mapping is a linear 
combination of $\tilde{e}$ shifted by a constant. 
We can  define the  class of programs we are interested in:

\begin{definition}
\label{definition:llm:tp}
Let $P$ be a flat CLP($\mathbb{C}$) program. We say that $P$ is
{\em alm-recurrent} if there exists an affine level mapping
$|\cdot |$ such that $P$ is recurrent with respect to it.
\end{definition}

\begin{example}
The CLP($\mathbb{Q}$) program $P$ from Example~\ref{example:72}
is alm-recurrent with respect to
$|p(x)| = 73 - x$. 
\end{example}

Clearly, if $P$ is alm-recurrent, then $P$ is recurrent thus terminating.
Let us show that alm-recurrence can be efficiently decided. We start
with proving this result for binary programs.

\begin{theorem}
\label{SVG:decidability}
Alm-recurrence of a binary constraint logic program $P$ over 
$\mathbb{Q}, \mathbb{Q}^+,\mathbb{R}$ and $\mathbb{R}^+$
is decidable in polynomial time with respect to the size of $P$.
\end{theorem}

\begin{proof} 
The proof is constructive: we provide a
decision procedure for alm-recurrence of binary constraint logic programs over 
$\mathbb{Q}, \mathbb{Q}^+,\mathbb{R}$ and $\mathbb{R}^+$. The decision procedure extends 
the algorithm proposed in~\cite{Sohn:van:Gelder} for termination of Prolog programs
(abstracted as CLP($\mathbb{N}$) programs) to binary CLP($\mathbb{C}$) where $\mathbb{C}$ is
$\mathbb{Q}, \mathbb{Q}^+,\mathbb{R}$ or $\mathbb{R}^+$.
The algorithm  tries to find
an affine level mapping showing that $P$ is alm-recurrent
by examining each user-defined predicate symbol $p$
of a binary CLP program  $P$ in turn 
(the precise order does not matter).
For every  rule $r$, say
$p(\tilde{x_p}) \leftarrow c,q(\tilde{x_q})$,
we test the satisfiability of $c$. For the domains we consider,
it can be done in polynomial time \cite{Khachiyan79}.
If $c$ is not satisfiable, we disregard this rule. Otherwise,
let $n_p$ and $n_q$ be the arities of $p$ and $q$.
For the rule $r$, recurrence is equivalent to:
\begin{equation}
\mathbb{C} \models c \rightarrow \left[ 
|p(\tilde{x_p})| \geq 1+|q(\tilde{x_q})| \land |q(\tilde{x_q})| \geq 0 \right]
\label{recc}
\end{equation}
Note that the condition
$c\rightarrow |p(\tilde{x_p})|\geq 0$ can be omitted as
it is implied by (\ref{recc}).
Formula (\ref{recc}) is logically equivalent 
to $\mathbb{C} \models c \rightarrow  |p(\tilde{x_p})| \geq 1+|q(\tilde{x_q})|$
and $\mathbb{C} \models c \rightarrow |q(\tilde{x_q})| \geq 0$.
Let $\tilde{x_p}$ be $(x_{p,1}, \ldots, x_{p, n_p})$, $\tilde{x_q}$ be $(x_{q,1}, \ldots, x_{q, n_q})$
and let $\mu_{p,0},\ldots,\mu_{p,n_p},\mu_{q,0},\ldots,\mu_{q,n_q}\in \mathbb{R}$ be such that 
for any atom $p(e_1,\ldots,e_{n_p})\in B_P^\mathbb{C}$ and any atom
$q(e_1,\ldots,e_{n_q})\in B_P^\mathbb{C}$:
$|p(e_1,\ldots,e_{n_p})| = \mu_{p,0} + \sum_{i=1}^{n_p} \mu_{p,i} e_i$
and 
$|q(e_1,\ldots,e_{n_q})| = \mu_{q,0} + \sum_{i=1}^{n_q} \mu_{q,i} e_i$.
Hence, $c$ should imply $(\mu_{p,0} - \mu_{q,0}) + \sum_{i=1}^{n_p} \mu_{p,i} x_{p,i} + \sum_{i=1}^{n_q} (-\mu_{q,i}) x_{q,i}\geq 1$ and
$\mu_{q,0} + \sum_{i=1}^{n_q} \mu_{q,i} x_{q,i}\geq 0$.
For the sake of uniformity, we rewrite the second inequality as
$\mu_{q,0} + \sum_{i=1}^{n_p} 0 x_{p,i} + 
\sum_{i=1}^{n_q} \mu_{q,i} x_{q,i}\geq 0$. Both inequalities can be 
presented using the scalar product notation as $\tilde{\mu}\tilde{x}\geq 1$
and $\tilde{\mu}'\tilde{x}\geq 0$, where: 
\[ \begin{array}{lll}
\tilde{x} & = & (x_0, x_{p,1}, \ldots, x_{p, n_p}, x_{q,1}, \ldots, x_{q, n_q}) \\
x_0       & & \mbox{\rm is a new variable fixed to $1$ and 
           used to obtain the free coefficient in the product}\\
\tilde{\mu} & = & (\mu_{p,0} - \mu_{q,0}, \mu_{p,1}, \ldots, \mu_{p,n_p}, -\mu_{q,1}, \ldots, -\mu_{q,n_q})\\
\tilde{\mu}' & = & ( \mu_{q,0},0, \ldots, 0, \mu_{q,1}, \ldots, \mu_{q,n_q}).
\end{array} \]


Hence, the binary rule $r$ gives rise to the  
following two \emph{pseudo}
linear programming problems. The problems are {\em pseudo} linear rather 
than linear
because \emph{symbolic} parameters appear in the objective functions.
\begin{equation}\label{eqprimal1}
minimise\ \ \theta = \tilde{\mu}\tilde{x} \ \  \mathit{subject \ to}    ~c \land x_0=1 \ \
\end{equation}
\begin{equation}\label{eqprimal2}
minimise\ \ \delta = \tilde{\mu}'\tilde{x}    \ \     \mathit{subject \ to}  ~c \land x_0=1
\end{equation}

We note that $c \land x_0=1$ is satisfiable as $c$ is satisfiable and $x_0$ 
is a new variable, and we rewrite $c \land x_0=1$ as $A \tilde{x} \geq b$ 
in the standard way~\cite{Schrijver86a}. 
An affine level mapping $|\cdot |$ ensuring recurrence  exists 
at least for this rule if and only if 
$\theta^* \geq 1$ and $\delta^* \geq 0$,
where $\theta^*$ and $\delta^*$
denote the minima of the corresponding objective functions. 
Because of the symbolic constants $\mu_{p,i}$ and $\mu_{q,i}$,
neither (\ref{eqprimal1}) nor (\ref{eqprimal2}) is a  linear programming problem. 
Now,  the idea is to consider the dual form:
\begin{equation}
maximise\ \ \eta= b^T  \tilde{y} \ \mathit{subject\ to} \  
A ^T \tilde{y} = \tilde{\mu}^T \land \tilde{y} \geq 0
\label{eqdual1}
\end{equation}
\begin{equation}
maximise\ \ \gamma= b^T  \tilde{z} \ \mathit{subject\ to} \  
A^T \tilde{z} = \tilde{\mu}'^T \land  \tilde{z} \geq 0
\label{eqdual2}
\end{equation}
where $\tilde{y}$ and $\tilde{z}$ are tuples of adequate length of new variables.
By the duality theorem of linear programming
which holds in $\mathbb{C}$
(see \cite{Schrijver86a} for instance), 
we have $\theta^* = \eta^*$ and $\delta ^*=\gamma^*$.
Furthermore, we observe that $\tilde{\mu}$ appears linearly in the dual problem (\ref{eqdual1}).
Hence the constraints of (\ref{eqdual1})
can be rewritten, by adding $\eta \geq 1$ 
as a set of linear inequations denoted $S_r^{p\geq 1+q}$.
Similarly,  the constraints of (\ref{eqdual2})
can be rewritten, by adding $\gamma \geq 0$ 
as a set of linear inequations, denoted $S_r^{q\geq 0}$.
Let us define  $\mathrm{defn}_P(p)$ as the set of binary rules defining $p$ in $P$,
$S_p$ as the conjunction $\bigwedge_{r \in \mathrm{defn}_P(p)}
[S_{r}^{p\geq 1+q} \land S_r^{q \geq 0}]$, and
$S_P$ as the conjunction $\bigwedge_{p \in \Pi_P} S_p$.
We have by construction $S_P$ is satisfiable
if and only if  there exists a  affine level mapping  ensuring recurrence 
of $P$.

Moreover, as $P$ is a finite set of binary rules,
computing $S_P$ can be done in polynomial time with respect to the size of $P$
and results in a constraint the size of which is also polynomial
with respect to the size of $P$.
Finally, testing satisfiability of $S_P$
in $\mathbb{Q}$, $\mathbb{Q}^+$, $\mathbb{R}$, and 
$\mathbb{R}^+$ can be done in polynomial time \cite{Khachiyan79}.
\end{proof}
\begin{example} 
Applying the algorithm to the example \ref{example:72},
we obtain the following two pseudo linear programming problems corresponding to
(\ref{eqprimal1}) and (\ref{eqprimal2}), respectively:
\begin{equation*}
minimise\ \ \theta = \mu_{p,1} x_1 - \mu_{p,1} x_2 \ \  \mathit{subject \ to}\;\;    72\geq x_1 \land x_2 = x_1 + 1 \land x_0 = 1 \ \
\end{equation*}
\begin{equation*}
minimise\ \ \delta = \mu_{p,0} + \mu_{p,1} x_2    \ \   \mathit{subject \ to}\;\;   72\geq x_1 \land x_2 = x_1 + 1 \land x_0 = 1
\end{equation*}
Rewriting the system of constraints as $A\tilde{x}\geq b$ and 
switching to the dual form, we get the system $S_P$:
$$
\left\{\begin{array}{l}
\eta=y_1-y_2-72*y_3+y_4-y_5, \\
\eta \geq 1, \\
y_1-y_2=0,-y_3-y_4+y_5=\mu_{p,1},\\
y_4-y_5= -\mu_{p,1}, \\
y_1\geq 0, \\
y_2\geq 0, \\
y_3\geq 0, \\
y_4\geq 0,\\
y_5\geq 0
\end{array}
\right\} \cup
\left\{
\begin{array}{l}
\gamma=z_1-z_2-72*z_3+z_4-z_5, \\
\gamma \geq  0, \\
z_1-z_2=\mu_{p,0},\\
-z_3-z_4+z_5=0,\\
z_4-z_5= \mu_{p,1},\\
z_1\geq 0,\\
z_2 \geq 0,\\
z_3 \geq 0,\\
z_4 \geq 0,\\
z_5 \geq 0
\end{array}
\right\}$$
Since $S_P$ is satisfiable, $P$ is alm-recurrent. Note that projecting
$S_P$ onto the $\mu_{p,i}$'s 
gives $\{\mu_{p,0}+73*\mu_{p,1} \geq 0, \mu_{p,1} \leq -1 \}$. Any solution
to this last constraint is a level mapping ensuring alm-recurrence of $P$.
\end{example}

An immediate consequence of the result above is that
recurrence with affine level mappings is also P-time decidable
for non-binary CLP($\mathbb{R}$) program with rules which 
contain more than one atom in their bodies. Formally, the following
theorem holds.

\begin{theorem}
\label{alm:dec}
Alm-recurrence of a constraint logic program $P$ over
$\mathbb{Q}, \mathbb{Q}^+,\mathbb{R}$ and $\mathbb{R}^+$
is decidable in polynomial time with respect to the size of $P$.
\end{theorem}
\begin{proof}
Let $P$ be a constraint logic program. Let $P'$ be the binary 
constraint logic program such that for every rule
$q_0(\tilde{y_0}) \leftarrow c, q_1(\tilde{y_1}), \ldots,q_n(\tilde{y_n})$ with $n \geq 1$
in $P$, $P'$ contains the following rules:
\begin{eqnarray*}
&& q_0(\tilde{y_0}) \leftarrow c, q_1(\tilde{y_1}).\\
&& \ldots\\
&& q_0(\tilde{y_0}) \leftarrow c, q_n(\tilde{y_n}).
\end{eqnarray*} 
and nothing else.
From Definition~\ref{recurrence}, we note that $P$ is recurrent if and only if $P'$ is recurrent.
Moreover, the size of $P'$ is polynomial in the size of $P$.
Hence, by Theorems~\ref{SVG:decidability},
alm-recurrence of $P'$ is P-time decidable.
\end{proof}

Although the technique above is not 
complete for programs over
$\mathbb{N}$, it is a sound way to prove recurrence 
of programs over this domain:
if a program is recurrent over $\mathbb{Q}$, it is also
recurrent over $\mathbb{N}$. 
For binary programs, as we allow 
negative coefficients in the level mapping,
we get a more powerful criterion
than the one proposed in \cite{Sohn:van:Gelder}. For instance, termination of
Example~\ref{example:72} (considered as a CLP($\mathbb{N}$) program) cannot
be proved by~\cite{Sohn:van:Gelder}.

For binary CLP($\mathbb{Q}$) programs,
the decision procedure described above has been 
{\em prototyped} in SICS\-tus Prolog~\cite{SICStus:Manual}
using the Simplex algorithm~\cite{Dantzig}
and a Fourier-based projection operator~\cite{Holzbaur95}
to ease manual verification. Therefore, the complexity
of the prototype is not polynomial.
The implementation is available at
{\small \url{http://www.univ-reunion.fr/~gcc/soft/binterm4q.tgz}}


\section{Related Works}
\label{Related Works}

The basic idea of identifying decidable and undecidable subsets 
of logic programs goes back to~\cite{Devienne:Lebegue:Routier}.

Recently, decidability of classes of imperative programs has been studied in
\cite{Cousot05-VMCAI,Podelski:Rybalchenko,Tiwari}. Tiwari considers 
real-valued 
programs with no nested loops and no branching inside a loop \cite{Tiwari}. 
Such programs correspond to one-binary-rule 
CLP($\mathbb{R}$). 
The author provides decidability 
results for subclasses of these programs. Our approach does not restrict nesting of loops and it allows internal {\em branching}. While in general
termination of such programs is undecidable~\cite{Tiwari}, we identified
a subclass of programs with decidable termination property.
Termination of the following CLP($\mathbb{R}$) program and its imperative equivalent can be shown by our method but not by the one proposed in~\cite{Tiwari}.
\begin{example}
\label{ex:disjunction}
\[ \begin{array}{lll}
q(x) & \leftarrow & -20\leq x, x\leq 20, y + 5 = x, q(y).\\
q(x) & \leftarrow & 0\leq x, x\leq 100, y + 1 = x, q(y).
\end{array} \]
\[
\begin{array}{l}
\mbox{\sl while}\;\;((-20 \leq  x \leq  20)\;\;\mbox{\sl or}\;\;(0 \leq  x \leq  100))\;\;\mbox{\sl do} \\
\hspace{0.5cm}   \mbox{\sl if}\;\;(-20 \leq  x \leq  20)\;\;x = x-5\;\;\mbox{\sl fi}\\
\hspace{0.5cm}   \mbox{\sl if}\;\;(0 \leq  x \leq  100)\;\; x = x-1\;\;\mbox{\sl fi}\\
\mbox{\sl od}
\end{array}
\]
\end{example}

Similarly to \cite{Tiwari}, Podelski and Rybalchenko \shortcite{Podelski:Rybalchenko}
 have considered programs
with no nested loops and no branching inside a loop. However, they focused on
integer programs and provide a polynomial time decidability technique for a 
subclass of such programs. In case of general programs their technique can be 
applied to provide a sufficient condition for liveness. 

In a recent paper, Cousot~\shortcite{Cousot05-VMCAI} applied abstraction techniques 
and langrangian relaxation to prove termination. Extension of the basic technique
should be able to analyse loops with disjunctions in their condition such as 
Example~\ref{ex:disjunction}. However, complexity of the approach is not discussed
and it is not clear whether the technique is complete for some class of programs.

One might like to investigate a more expressive language of constraints
including polynomials. Recall that we require the constraints domain to be 
{\em ideal}, i.e., one needs a decision procedure for existentially closed
conjunctions. Such a decision procedure exists, for instance, for real-closed 
fields such as ${\mathbb R}$~\cite{Tarski:fund:math,Renegar}.
For some domains such as ${\mathbb Q}$, existence of a decision procedure
is still an open problem, although it seems to be unlikely~\cite{Pheidas}.
If one restricts attention to real-closed fields, one might even consider
polynomial level-mappings of a certain power rather than the affine ones. 
One can show that in this case proving recurrence 
is equivalent to determining satisfiability of 
the equivalent quantifier-free formula
\cite{Tarski:fund:math,Tarski:decision:method}. Hence, recurrence is still
decidable in this case. Although the known complexity bound
of determining the equivalent quantifier-free formula 
given an existential formula is a double
exponential~\cite{Basu:Pollack:Roy:JACM,Collins},
to the best of our knowledge
the complexity of the subclass of formulas which  we obtain 
is an open question.


\section{Conclusion}
\label{Conclusion}
In this paper we have considered constraints solving over the rationals and
the reals. For these domains we have identified a class of CLP programs 
such that an affine level mapping is sufficient to prove their recurrence. 
We have seen that membership to this class is decidable and presented
a polynomial-time decision procedure. The decision procedure can also be used
as a sound termination proof technique for binary CLP(${\mathbb N}$)
and has been prototyped in SICStus Prolog for binary CLP(${\mathbb Q}$).

\section*{Acknowledgements} We thank
the referees for useful  suggestions.

\bibliographystyle{acmtrans}
\bibliography{paper}

\end{document}